\documentclass[aps,prl,nobibnotes,twocolumn,superscriptaddress,bibliography]{revtex4-2}

\usepackage{amsfonts}
\usepackage{mathrsfs}
\usepackage{amsmath}
\usepackage{color}
\usepackage{graphicx}
\usepackage{bm}
\usepackage{amssymb}
\usepackage{xspace}
\usepackage{epstopdf}
\usepackage{dcolumn}
\usepackage{longtable}
\usepackage{multirow}
\usepackage{float}
\usepackage{comment}

\def\eg{{\it e.g.}\ }
\def\ea{{\it et al.}}

\usepackage[colorlinks=true, letterpaper=true, pdfstartview=FitV, linkcolor=blue, citecolor=blue, urlcolor=blue]{hyperref}

\makeatother

\begin{document}
\title{From Atomic Semimetal to Topological Nontrivial Insulator}
\author{Xiao-Ping Li}
\affiliation{Centre for Quantum Physics, Key Laboratory of Advanced Optoelectronic Quantum Architecture and Measurement (MOE), School of Physics, Beijing Institute of Technology, Beijing, 100081, China}
\affiliation{Beijing Key Lab of Nanophotonics \& Ultrafine Optoelectronic Systems, School of Physics, Beijing Institute of Technology, Beijing, 100081, China}
\author{Da-Shuai Ma}
\email{madason.xin@gmail.com}
\affiliation{Centre for Quantum Physics, Key Laboratory of Advanced Optoelectronic Quantum Architecture and Measurement (MOE), School of Physics, Beijing Institute of Technology, Beijing, 100081, China}
\affiliation{Beijing Key Lab of Nanophotonics \& Ultrafine Optoelectronic Systems, School of Physics, Beijing Institute of Technology, Beijing, 100081, China}
\author{Cheng-Cheng Liu}
\affiliation{Centre for Quantum Physics, Key Laboratory of Advanced Optoelectronic Quantum Architecture and Measurement (MOE), School of Physics, Beijing Institute of Technology, Beijing, 100081, China}
\affiliation{Beijing Key Lab of Nanophotonics \& Ultrafine Optoelectronic Systems, School of Physics, Beijing Institute of Technology, Beijing, 100081, China}
\author{Zhi-Ming Yu}
\affiliation{Centre for Quantum Physics, Key Laboratory of Advanced Optoelectronic Quantum Architecture and Measurement (MOE), School of Physics, Beijing Institute of Technology, Beijing, 100081, China}
\affiliation{Beijing Key Lab of Nanophotonics \& Ultrafine Optoelectronic Systems, School of Physics, Beijing Institute of Technology, Beijing, 100081, China}
\author{Yugui Yao}
\email{ygyao@bit.edu.cn}
\affiliation{Centre for Quantum Physics, Key Laboratory of Advanced Optoelectronic Quantum Architecture and Measurement (MOE), School of Physics, Beijing Institute of Technology, Beijing, 100081, China}
\affiliation{Beijing Key Lab of Nanophotonics \& Ultrafine Optoelectronic Systems, School of Physics, Beijing Institute of Technology, Beijing, 100081, China}

\begin{abstract}
Topological band insulators and (semi-) metals can arise out of atomic insulators when the hopping strength between electrons increases. 
Such topological phases are separated from the atomic insulator by a bulk gap closing. 
In this work, we show that in many (magnetic) space groups, the crystals with certain Wyckoff positions and orbitals being occupied must be semimetal or metals in the atomic limit, \eg the hopping strength between electrons is infinite weak but not vanishing, which then are termed atomic (semi-)metals (ASMs). 
We derive a sufficient condition for realizing ASMs in spinless and spinful systems. 
Remarkably, we find that increasing the hopping strength between electrons may transform an ASM into an insulator with both symmetries and electron fillings of crystal are preserved. 
The induced insulators inevitably are topologically non-trivial and at least are obstructed atomic insulators (OAIs) that are labeled as trivial insulator in  topological quantum chemistry website.
Particularly, using silicon as an example, we show ASM criterion can discover the OAIs missed by the recently proposed criterion of filling enforced OAI. 
Our work not only establishes an efficient way to identify and design non-trivial insulators but also predicts that  the group-IV elemental semiconductors are ideal candidate materials for OAI.
\end{abstract}
\maketitle

\textit{\textcolor{blue}{Introduction}}\textit{.} 
The past decade has witnessed the prosperity and development of the band topology in condensed matter physics \cite{zhang2011RMP,Kane2010RMP,Chiu2016RMP}. 
The topology in electronic bands has been recognized for a long time, such as quantum anomalous Hall effect \cite{Haldane1988}. 
However, in the early stage the topological phases in crystals were considered very rare and required strict external conditions \cite{bernevig2013topological}. 
The discovery of topological insulator and topological Weyl semimetal open up a new direction for realizing topological phases in noninteracting systems by highlighting spin-orbit coupling (SOC) effect \cite{KaneMele2005,BHZ2006,wan2011weyl}. 
Many materials constructed by heavy atoms are predicted as topological materials with nontrivial boundary states \cite{fangQAH2010,yaoQAH2010,Bi2Se32009,2DBi2011,BiBr2014,SnTe2012,weng2015weyl,HgCrSe,CoSnS,Na3Bi2012,Cd3As2,yuHONL,type-II-Weyl,shi2021chargeWeyl,Type-III-Weyl}. 
However, compared with the realistic materials in database, the number of the topological materials is still very small.

The breakthrough comes from the establishment of two equivalent theories: topological quantum chemistry (TQC) \cite{TQC,MTQC}, symmetry indicator theory \cite{PoSI,MpoSI,TCIBC}, and related theories \cite{songSI,MSI}, both of which present a quantitative description of the atomic insulators respectively based on the elementary band representations (EBRs) and the symmetry eigenvalues of the bands at high-symmetry points. 
The application of these two theories to the material database predicted that thousands of realistic materials are topologically nontrivial \cite{wangTM,wanTM,fangTM,xuTM}. 
However, there is a caveat for the classification, due to the definition of topologically trivial insulator. 
In the definition, the topologically trivial insulator is the material that its band representation (BR) of valence bands can be represented by a sum of EBRs \cite{TQC}, which means the trivial state can be described by a set of exponentially localized Wannier functions \cite{WFO}. 
For atomic insulators, the BRs are induced from the orbitals locating at occupied Wyckoff positions. 
However, for certain materials, the BRs also can be induced from orbitals locating at the unoccupied Wyckoff positions.
Such materials are then termed obstructed atomic insulators (OAIs) to distinguish them from the atomic insulators \cite{TQC,WFO,EBR,song2018prx,wang2021}. 
For now, OAI has only been proposed in several theoretical models and a few artificial systems \cite{lightOAI,DiracOAI}. Very recently, Xu \ea  \cite{xu2021filling} developed an efficient criterion to identify filling-enforced OAIs without calculating the representation of band structures, and found that the OAI can appear in a large number of crystal materials. 

Meanwhile, the group-IV elemental semiconductors, such as silicon, diamond, and germanium, play a central role in the modern microelectronics industry. 
These elementary materials are widely studied and were considered as topologically trivial semiconductors with indirect band gap, due to negligible SOC and sizable band gap. 
According to both the  \href{https://www.topologicalquantumchemistry.com}{database of TQC website} and filling-enforced OAI~\cite{xu2021filling}, the group-IV elemental semiconductors indeed are classified into topologically trivial insulators. 
However, a recent study shows the silicon has topological surface state \cite{Si001,Si100,Murakami2020}, similar to the OAI. 
Thus, several natural questions arise: does the silicon and the other group-IV
elemental semiconductors are OAI? 
If they are, then does there exist an efficient method to identify these OAIs missed by the criterion of filling-enforced OAIs?

\begin{figure}[t]
\includegraphics[width=8.2cm]{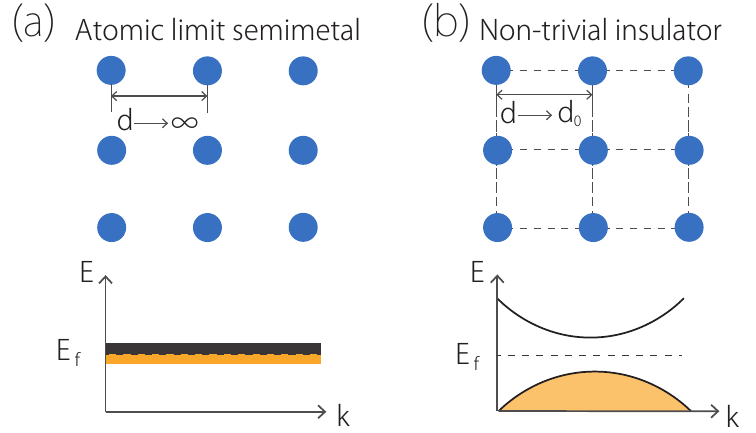} \caption{(a)-(b) Schematic of the crystal structure in atomic limit and crystal phase. The occupied bands are highlighted in claybank. See Fig. \ref{fig3} for  silicon as a specific example. \label{fig1}}
\end{figure}

In this work, we address these questions in the affirmative.
We predict an interesting and remarkable metal-insulator transition driven by the hopping strength between electrons. 
For many crystals with certain Wyckoff positions and orbitals being occupied, they must be (semi-)metal when the hopping strength between electrons is infinite weak but not vanishing, due to the presence of partially filled EBRs (see Fig.~\ref{fig1}(a)). 
By increasing the hopping strength, the systems can be transformed into insulators due to band inversion when both symmetries and electron fillings of crystal are preserved, as illustrated in Fig.~\ref{fig1}(b). 
The insulators transformed from atomic semimetal must be topologically nontrivial. 
Since the stable and fragile topological insulator generally can be diagnosed by TQC and symmetry indicator theory, here we focus on the OAI. 
In other words, we can apply the condition of atomic semimetal to the materials classified as trivial insulators in the database of TQC website and Ref.~\cite{xu2021filling},  and the materials that satisfy the condition of atomic semimetal should be OAIs. 
We also present a sufficient condition for the atomic semimetals in spinless and spinful systems, which only depend on the crystal structure and the orbits of the atoms. 
Thus using the condition of atomic semimetal to diagnose OAI will be very efficient, as it does not resort to the time-consuming DFT calculations. 
By applying the criterion to the group-IV elemental semiconductors, we find they fall into the condition of atomic semimetal, indicating that they would be OAI rather than a trivial semiconductor. 
This is further confirmed by DFT calculations.
Particularly, the condition of atomic semimetal also offers useful guidance for designing topological states with light atoms, as in such case no matter whether the system is semimetal or an insulator, it must be topologically nontrivial. 

\textit{\textcolor{blue}{Condition of Atomic semimetal}}\textit{.}
We consider a three-dimensional (3D) material system belonging to (magnetic) space group $\boldsymbol{G}$. 
The atoms in this system can be divided into $N$ non-equivalent sets, and the $i$-th set of atoms can be labeled as $\{i;w_{i},\sum_{l}\ ^{n_{l}}O_{l}\}$ where $w$ denotes the Wyckoff position occupied by the atoms, $O_{l}$ is the atomic orbit of the atom with (total) angular momentum $l$, and $n_{l}$ is the number of electrons residing at the corresponding orbit. 
For spinless systems, $l=0,1,2...$ is an integer and for spinful systems, $l=\frac{1}{2},\frac{3}{2}...$ is a half odd integer. 
Assuming the site symmetry of $w_{i}$ Wyckoff position is $\boldsymbol{g}_{w_{i}}$, and the irreducible representations (IRRs) of $\boldsymbol{g}_{w_{i}}$ are $\rho_{w_{i}}^{m=1,2...}$. 
The atomic orbit $O_{l}$ of the atom at $w_{i}$ generally would be splitted into different IRRs ($\rho_{w_{i},l}^{m}$) of $\boldsymbol{g}_{w_{i}}$ by the crystalline field with symmetry of $\boldsymbol{g}_{w_{i}}$. 
For example, in spinless systems the five $l=2$ ($d$) orbits would be two levels, represented two IRRs $E$ and $T_{2}$ by the crystalline field with symmetry of $T_{d}$ point group, as listed in Table.~\ref{tab1}. 
{\renewcommand\arraystretch{1.3}
\begin{table}[b]
\caption{The atomic orbitals and IRRs of $T_{d}$ point group. The third column are the BRs induced by the corresponding atomic orbitals in $8a$  Wyckoff position of 227 space group.}\label{tab1}
\begin{ruledtabular} %
\begin{tabular}{lll}
Orbital & IRRs & BRs \\
		 \hline
		$s$ ($l$=0) & $A_{1}$ & $A_{1}\uparrow\boldsymbol{G}(2)$  \\
		$p$ ($l$=1) & $T_{2}$ & $T_{2}\uparrow\boldsymbol{G}(6)$  \\
		$d$ ($l$=2) & $E\oplus T_{2}$ & $E\uparrow\boldsymbol{G}(4)\oplus T_{2}\uparrow\boldsymbol{G}(6)$  \\
        $f$ ($l$=3) & $A_{1}\oplus T_{1}\oplus T_{2}$ & $A_{1}\uparrow\boldsymbol{G}(2)\oplus T_{1}\uparrow\boldsymbol{G}(6)\oplus T_{2}\uparrow\boldsymbol{G}(6)$  \\     
\end{tabular}\end{ruledtabular}
\end{table}
}
According to TQC, the IRRs $\rho_{w_{i},l}^{m}$ in space group $\boldsymbol{G}$ can be expressed as $\rho_{w_{i},l}^{m}\uparrow\boldsymbol{G}$, for which the dimension is denoted as $d(\rho_{w_{i},l}^{m}\uparrow\boldsymbol{G})$.
Then a sufficient condition for a system to be atomic semimetal phase is equivalent to there does not exist a solution for the following equation
\begin{eqnarray}
N_{occ} & = & \chi\sum_{i,l,m}c_{i,l,m}d(\rho_{w_{i},l}^{m}\uparrow\boldsymbol{G}),\ \ c_{i,l,m}\in\{0,1\},\label{eq:AM}
\end{eqnarray}
where $N_{occ}$ is the total number of the electrons of the material and $\chi=2$ for spinless systems and $\chi=1$ for spinful systems.
Notice that the summation in Eq.~(\ref{eq:AM}) is constrained in the occupied Wyckoff position and the occupied atomic orbits. 
Besides, while the ASM also is filling enforced, it can not be fully captured by the theory proposed by H. Watanabe \ea ~\cite{watanabeFE2016}, as the ASM predicted here can be transformed into a gapped state with maintaining the electron fillings, as shown in the following discussion of silicon. 
Particularly, the criterion of ASM is quite different from that of filling enforced OAI where the occupied Wyckoff positions always have higher multiplicity than that of a $necessary$ empty Wyckoff position.  Thus, applying the criterion of ASM to the trivial insulators in TQC database can discover the OAIs that can not be diagnosed by the criterion of filling enforced OAI.
The proposal of ASM is fundamentally interesting in that it shows the  hopping strength of electrons can transform semimetal phase into topologically nontrivial insulator phase. 
This also indicates that all the material systems fall into the ASM criterion would be topologically nontrivial, as they inevitably can not be adiabatically transformed into atomic insulator without a gap closing or opening. 

\begin{figure}[t]
\includegraphics[width=8.2cm]{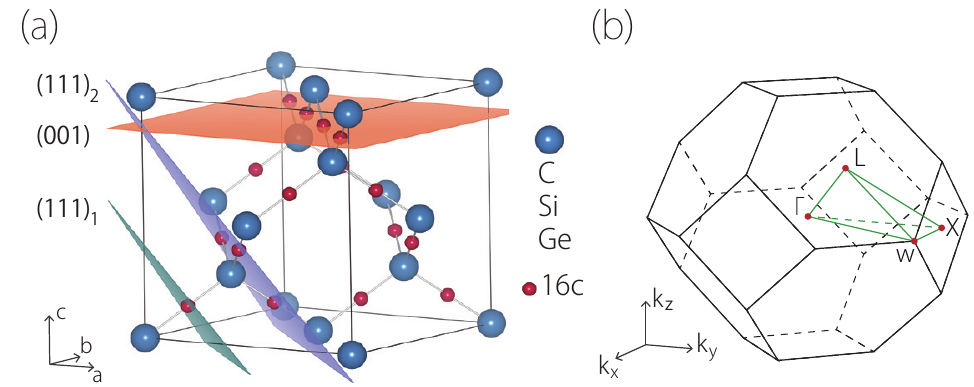} \caption{(a) Crystal structure of group-IV elemental semiconductor materials. The Wannier charge centers are denoted by small red ball located at Wyckoff position 16$c$. The three kinds of surface termination \eg $(111)_{1}$, (001), and $(111)_{2}$  are highlighted in green, blue and orange, respectively. (b) Bulk Brillouin zone of group-IV elemental semiconductor materials.
\label{fig2}}
\end{figure}
\begin{figure}[t]
\includegraphics[width=8.8cm]{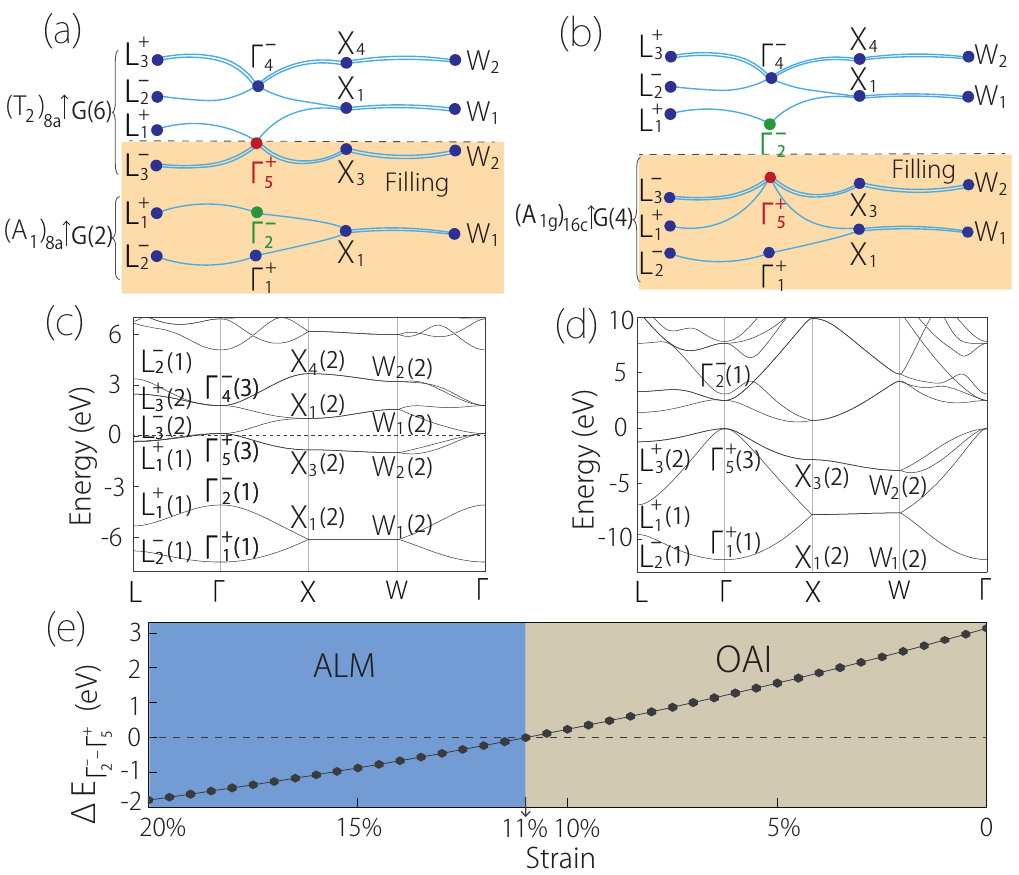} \caption{(a) Schematic of the EBRs $(A_{1})_{8a}\uparrow G(2)$ and $(T_{2})_{8a}\uparrow G(6)$ induced from $s$ and $p$ orbitals at  Wyckoff  position 8$a$ with 8 electrons (4 bands in spinless case). (b) Schematic of the EBRs that undergoing a band inversion between $\Gamma_{2}^{-}(1)$ and $\Gamma_{5}^{+}(3)$. (c) The band structure of silicon under $40\%$ tensile stress. (d) The band structure of silicon with the fully optimized lattice constant. The BRs of bands at the high symmetry points are also given. The superscript $+/-$ denotes the parity. (e) The  phase diagram and  band gap between $\Gamma_{2}^{-}(1)$ and $\Gamma_{5}^{+}(3)$ as a function of the lattice constant.
 \label{fig3}}
\end{figure}

\textit{\textcolor{blue}{Silicon and ASM-insulator transition}}\textit{.} 
We take silicon (Si) as an example to check the ASM criterion and study the hopping strength-driven ASM-insulator transition. 
Since the SOC effect of Si is negligible, the bulk silicon can be considered as a spinless system. 
Moreover, the bulk silicon is well known to have tetrahedral crystal structure with space group Fd3m (No.~227), as shown in Fig.~\ref{fig2}(a). 
Each unit cell of elemental Si contains 8 Si atoms with perfect tetrahedral $sp3$ bonding residing at the Wyckoff position $8a$, and there are 2 Si atoms in each primitive unit cell. 
Thus, the multiplicity of Wyckoff position $8a$ whose site symmetry group is $T_{d}$ point group is 2 for primitive unit cell. 
The arrangement of electrons of the Si atom in outer shell is $3s^{2}3p^{2}$, indicating the total number of electrons in a primitive unit cell of silicon is $8$. 
According to the crystalline field splitting, the IRRs from the $s$ and $p$ orbits at $8a$  are $A_{1}$ and $T_{2}$ (\eg $\rho_{8a,l=0}^{1}=A_{1}\uparrow\boldsymbol{G}$ and $\rho_{8a,l=1}^{5}=T_{2}\uparrow\boldsymbol{G}$), respectively. From the \href{https://www.cryst.ehu.es/}{Bilbao Crystallographic Server}, one knows the dimension of the two IRRs induced from $A_{1}$ and $T_{2}$ orbitals are $d(A_{1}\uparrow\boldsymbol{G})=2$ and $d(T_{2}\uparrow\boldsymbol{G})=6$. According to Eq.~(\ref{eq:AM}) , we immediately have
\begin{eqnarray}
\nexists\ c_{i,l,m}\in\{0,1\},\  & s.t.\  & 8=2\times(2c_{1,0,1}+6c_{1,1,2}),
\end{eqnarray}
showing that in the atoms limit the silicon must be a semimetal, as illustrated in Figs.~\ref{fig3}(a) and (c). 

However, as we know the pristine silicon is a semiconductor, and then there would exist an interaction-driven ASM-insulator transition in silicon. 
To directly demonstrate it, we calculate the electronic properties of silicon under different tensile stresses. 
The results and the obtained phase diagram are shown in Fig.~\ref{fig3}(e). 
Without stress, the silicon is calculated as a semiconductor, as it should be. 
By applying stress, the band gap at $\Gamma$ point will decrease and finally closes under $\sim11\%$ tensile stress. 
The band inversion occurring at $\Gamma$ leads to a semimetal state for the silicon. 
Keep increasing tensile, there does not exist another phase transition, namely, the silicon under atomic limit is a semimetal state, consistent with the above analysis. 
Moreover, by calculating the BRs of the silicon under $40\%$ stress, we find that the BR of the lower eight bands (four valence bands and four conduction bands) indeed are induced by the $s$ and $p$ orbits at $8a$ position, and a clear diagram of band inversion are shown in Figs.~\ref{fig3}(a) and (b).

\textit{\textcolor{blue}{OAI and floating surface states}}\textit{.} Since the semiconductor state of silicon is transformed from ASM, the pristine silicon would be topologically nontrivial. 
According to the database of TQC and Ref. \cite{xu2021filling}, silicon does not have stable and fragile topology. 
Thus one can expect that the silicon would be an OAI. 
To confirm it, we calculate the BRs of the valence bands of pristine silicon (see Fig.~\ref{fig3}(d) and details in Supplementary Materials~\cite{refsm} ) and find the BRs are solely induced by the $A_{1g}$ orbital in $D_{3d}$ point group at Wyckoff position $16c$ of space group Fd3m, which is an unoccupied position in silicon. 
In Fig.~\ref{fig2}(a), we also marked the $16c$ Wyckoff position with red and small balls. 
In other words, there is obstructed Wannier charge centers locating at the Wyckoff position $16c$. 
Hence, the result directly demonstrates that silicon is an OAI.

\begin{figure}[t]
\includegraphics[width=8.8cm]{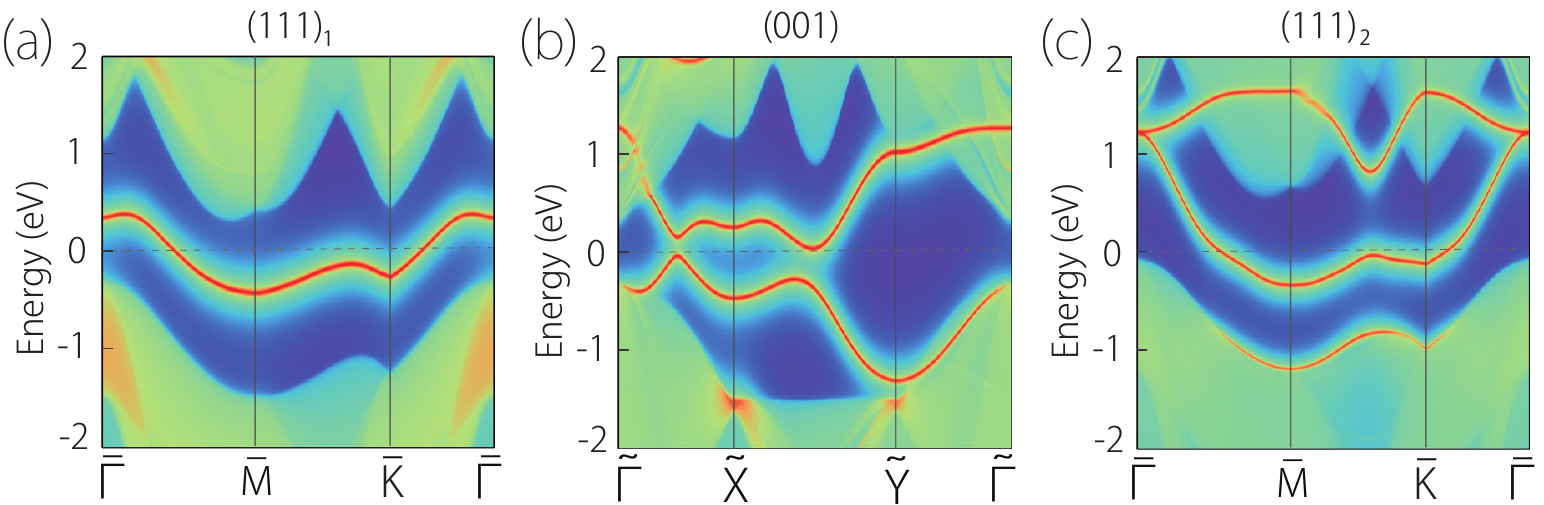} \caption{(a)-(c) Surface
spectra on $(111)_{1}$, $(001)$, and $(111)_{2}$ surfaces. The surface Brillouin zone and paths of surface bands are provided in Supplemental Material \cite{refsm}.
\label{fig4}}
\end{figure}

We then study the surface state of silicon. 
Distinguished from the Chern insulator and topological insulator, where the surface states always cross the bulk gap regardless of the boundaries, the surface states of OAI are floating bands appearing in the bulk gap and only occur on the boundaries that cut through the unoccupied Wyckoff position where the Wannier charge centers is locating at. 
This means that while the surface state of OAI is not as robust as that in Chern insulator and topological insulator, but its form can be more rich and varied. 
Interestingly, when we chose the three crystal planes labeled as $(111)_{1}$, $(001)$ and $(111)_{2}$ planes in Fig.~\ref{fig2}(a), as cleavage terminations, there would exist three completely different surface states on the corresponding boundaries. 
The calculated surface states of silicon for the three boundaries are plotted in Figs.~\ref{fig4}(a)-(c). 
One finds that $(111)_{1}$, $(001)$ and $(111)_{2}$ boundaries exhibit $1$, $2$ and $3$ floating surface bands, respectively. 
This is because that the three boundaries cut $1$, $2$ and 3 obstructed Wannier charge centers (see Fig.~\ref{fig2}(a)). 

Since the mismatch between the obstructed Wannier charge centers and atom's position can lead to a bulk electric polarization, the floating surface states can also be understood in terms of the Zak phase along a straight line normal to the boundary. 
The expression of Zak phase generally can be expressed as \cite{zakphase}
\begin{eqnarray}
\mathcal{Z}(\boldsymbol{k}_{\parallel}) & = & -i\sum_{n}\int_{0}^{2\pi}\left\langle u_{n\boldsymbol{k}}\left|\frac{\partial}{\partial k_{\perp}}\right|u_{n\boldsymbol{k}}\right\rangle dk_{\perp},
\end{eqnarray}
where the summation is performed on the occupied bands, $\boldsymbol{k}_{\parallel}$ ($\boldsymbol{k}_{\perp}$) denotes the momentum parallel (normal) to the boundary and $\left|u_{n\boldsymbol{k}}\right\rangle $ is the lattice periodic part of the Bloch wave function. 
For the Zak phases $\mathcal{Z}(\boldsymbol{k}_{\parallel})$ of a straight line normal to $(111)_{1}$ and $(111)_{2}$  boundaries, it is calculated as $\pi$ for any $\boldsymbol{k}_{\parallel}$, showing the nontrivial properties of silicon. 
However, the Zak phase $\mathcal{Z}(\boldsymbol{k}_{\parallel})$ of a straight line normal to $(001)$ boundary is calculated as $0$ for any $\boldsymbol{k}_{\parallel}$. 
The inconsistent between the trivial Zak phase and two floating surface bands on $(001)$ boundary is due to the fact the Zak phase is a $\mathbb{Z}_{2}$ topological quantum number and cannot capture the topological nature of systems with even floating surface bands. 
Therefore, for the prediction of the floating surface bands of OAIs, the obstructed Wannier charge centers calculated by TQC are more advantageous and can provide more accurate information.

\textit{\textcolor{blue}{Discussion and Conclusion}}\textit{.} The group-IV elemental materials, such as diamond, silicon, germanium and tin, have same crystal structure and similar arrangement of electrons in outer shell, \eg $ns^{2}np^{2}$ with $n=2,3,4,5$ for C, Si, Ge and Sn atoms, respectively. 
Hence, all the group-IV elemental materials satisfy the criterion of ASM and are topologically nontrivial, except that group-IV elemental semiconductors, \eg diamond, silicon and germanium are OAI, while Sn naturally is in the ASM phase \cite{refsm}. 

It should be noticed that Eq.~(\ref{eq:AM}) is a sufficient but not a necessary condition for ASM. 
To obtain a sufficient and necessary condition for ASM, one should have the information of the energy level of all the induced BRs, which can not be inferred from the crystalline structure and the atomic orbits. 
One possible way to find out all the topological materials satisfying the ASM criterion is to calculate the band structure of each realistic material in the database under sufficiently large tensile stress and check whether it is gapless or not. 
Besides, since there does not exist a sufficient and necessary condition for diagnosing all the possible topological materials, comparing the BRs induced by the valence band of each realistic material with a sufficiently large tensile stress and the BRs of the corresponding material without stress may be a possible effective method to address this task.

In summary, we propose the concept of ASM and establish a sufficient condition for ASM. 
We find group-IV elemental semiconductor: silicon satisfies the ASM criterion and then would be topologically nontrivial. 
Further calculations show silicon is an OAI. 
Based on the location of obstructed Wannier charge centers, we also discuss the floating surface states for three different boundaries of silicon, which presents a quantitative and physical description of the surface states of silicon. 
Due to mature theology, silicon would be the most ideal material for studying much more novel phenomena associate with OAI.
Moreover, the criterion of ASM is a very powerful tool that can not only identify topological nontrivial phases but also be used to design topological materials by placing specific atoms at Wyckoff positions in a given space group.

\acknowledgements
The authors thank J. Xun for helpful discussions. This work is supported by the National Key R\&D Program of China (Grants No.~2020YFA0308800), the NSF of China (Grants No.~11734003, No.~12061131002 and No.~12004035), the Strategic Priority Research Program of the Chinese Academy of Sciences (Grant No.~XDB30000000), and the Beijing Institute of Technology Research Fund Program for Young Scholars.
\bibliography{ref.bib}

\end{document}